\documentclass{aa}  
\usepackage{graphicx}
\usepackage[colorlinks=true,citecolor=blue]{hyperref} 
\usepackage{gensymb}
\usepackage{txfonts}
\usepackage{siunitx} % use for \si
\usepackage{ulem}    % use for \ce molecular notation
\usepackage{orcidlink} 
\usepackage[version=4]{mhchem}
   \title{Gas chemistry in the dust depleted inner regions \\
   of protoplanetary disks}

   \subtitle{I. Near-IR spectra and overtones }

   \author{J. Bethlehem \orcidlink{0009-0001-1761-7862}\inst{1,2},
          Ch. Rab \orcidlink{0000-0003-1817-6576} \inst{1,}\inst{3},
          I. Kamp \orcidlink{0000-0001-7455-5349}\inst{2},
          M. Flock \orcidlink{0000-0002-9298-3029}\inst{4},
          G. Bourdarot \orcidlink{0000-0002-6777-6386} \inst{1},
          P. Caselli \orcidlink{0000-0003-1481-7911}\inst{1}}

   \institute{Max-Planck Institute for extraterrestrial physics (MPE), Giessenbachstr. 1, 85748, Garching, Germany \email{J.bethlehem@rug.nl}
    \and
    Kapteyn Astonomical Institute, University of Groningen,
    P.O. Box 800, 9700 AV Groningen, The Netherlands
    \and
    University Observatory, Faculty of Physics, Ludwig-Maximilians-University at Munchen, Scheinerstr. 1, 81679 Munich, Germany
    \and
    Max-Planck-Institut fur Astronomie, Konigstuhl 17, D-69117 Heidelberg, Germany
    }

   \date{Received: 19 December 2025; accepted: 5 March 2026}
 
  \abstract
   {The molecular composition inside the dust sublimation zones of protoplanetary disks is mostly unknown but important to understanding terrestrial planet formation. A few molecules have been observed from this region, specifically \ce{CO}, \ce{H2O}, \ce{OH} and \ce{SiO}. The small surface area makes observing this region difficult, hence modeling is required to disentangle the innermost disk from regions further out.}
   {We model a protoplanetary disk around a Herbig-type star including the dust depleted inner region ($\approx$ 0.1-0.3 au) and aim to investigate the chemistry of this region and explain existing and future observations.}
   {We post-process the dust and gas distribution of a magnetohydrostatic model with the radiation thermochemical code ProDiMo to study the chemistry and to produce observables.}
   {We find that the dust free inner disk is a molecular rich environment, where besides \ce{CO} we also find \ce{H2}, \ce{H2O} and \ce{SiO}. The gas temperature profile is complex and fluctuates between 700 and 2000 K, which is warm enough to produce CO overtone line emission. Next to the CO overtone lines we also find strong high J-level fundamental CO lines between 4.3 and 4.6 \si{\mu m}. The elemental enrichment of \ce{Si} due to dust sublimation leads to 2 orders of magnitude more \ce{SiO} abundance. The \ce{SiO} gas has average temperatures of $\approx$ 1000 K resulting in strong \ce{SiO} overtone emission in the spectral range between 4 and 4.3 \si{\mu m}.}
   {We predict that the gas density in the dust depleted inner disk is high enough to allow for \ce{H2} formation, resulting in an molecular rich environment. For our representative Herbig model, the dust-depleted inner disk is responsible for at least 90\% of the line emission for \ce{CO} and \ce{H2O} between 1 and 28 \si{\mu m}. Next to CO overtone lines, \ce{SiO} overtone lines are expected to be an important tracer of a dust free inner disk.}
   
    \titlerunning{Hot chemistry from the dust depleted inner disk}
    \authorrunning{J. Bethlehem et al.}
   \keywords{Protoplanetary disks --
                dust sublimation --
                astrochemistry 
               }
\begin{document} 
   
\maketitle

%-------------------------------------------------------------------

\section{Introduction}\label{introduction}

The inner regions of protoplanetary disks (within $\approx$1 au) are important for the understanding of planet formation, as they mark the zone where the material originates that contributes to terrestrial planets. With the advent of GRAVITY on the Very Large Telescope Interferometer (VLTI) \citep{Gravity_first_light_2017} CO overtone line emission from the innermost part at sub au scales of disks can be spatially resolved. We aim to investigate the chemistry of the dust sublimation region, understanding the CO overtone line emission using a chemical/physical model of the dust depleted inner disk.

Close to the star, the gas and dust temperatures in the disk exceed the dust sublimation temperature (1300-2200 K). The transition zone, where dust grains sublimate is called the "dust inner rim" \citep{Dullemond_2010}. This inner rim was speculated to exist by \citet{Natta_2001} and \citet{Tuthill_2001}. In order to give a correct fit to the spectral energy distribution (SED) they invoked a puffed up hot dusty wall. \citet{Dullemond_2001} worked out the theory for a puffed-up inner rim, a vertical wall-like structure at the dust sublimation radius ($\sim$0.1 au), where grains are heated directly by stellar radiation and sublimate. This inner rim, also known as the dust sublimation radius, is not the same as the accretion/truncation radius. Inwards of this rim, dust cannot exist in thermal equilibrium such that this part of the disk is often called dust free or gaseous inner disk. The gaseous inner disk can extend inward to the truncation radius at $\sim$0.03--0.08 au \citep{Bouvier_2007}. In classical T~Tauri stars, this region is magnetically truncated by strong stellar dipole fields, which disrupt the inner disk and allow gas to accrete along magnetic field lines onto the star \citep{Koenigl_1991, Hartmann_1994, Bessolaz_2008, GRAVITY_2023}. Seen from the star, there is the truncation radius, then the co-rotation radius followed by the dust sublimation radius. In this paper, we will use the term 'dust depleted inner disk' for the region between the truncation radius and the dust sublimation radius \citep{gravity_2021b}.

The inner few au of a protoplanetary disk have been studied through spectroscopy of molecular and atomic lines in the near-and mid-infrared for both T~Tauri and Herbig type objects. \ce{CO}, \ce{H2O}, \ce{OH}, and hydrogen recombination lines (Br $\gamma$) have been observed \citep[e.g.][]{Najita_2003, Carr_2008, Thi_2005, Gravity_2021a, Banzatti_2015}. These observations indicate decoupled dust and gas temperature \citep{Salyk_2008}. The first recognized emission to probe the dust depleted inner disk is the CO overtone emission \citep[$\Delta \nu=2,\ \lambda=2.3\ \mu m$, e.g.][]{Najita_2007}. In a fraction of objects that show \ce{CO} overtone emission, also \ce{H2O} emission has been detected \citep{Najita_2000, Carr_2004, Thi_2005} and OH is observed in actively accreting sources that have \ce{CO} overtone and \ce{H2O} emission \citep{Najita_2007, Carr_2008}. More recently spectro-astrometric observations with VLT/CRIRES have revealed fundamental ro-vibrational \ce{CO} emission originating from both an outer and inner disk for a range of targets including T~Tauri and Herbig up to three solar mass based on the line broadening of \ce{CO} emission. Outer and inner here are two components and do not necessarily correspond to inside/outside the dust sublimation radius. The inner broad component originating from 0.05-3~au at 50-200 km/s while the outer narrow component traces distances of 0.1-10 au with velocities of 10-50 km/s \citep{Pontoppidan_2008, Banzatti_2015}. Furthermore, mid-infrared observations with JWST/MIRI have detected multiple gas-phase species in disks with large dust cavities. The spatial origin of the gas is based on line broadening caused by Keplerian rotation. In the T~Tauri system PDS 70, water emission was found at $\approx$ 0.1~au \citep{Perotti_2023} and in the T~Tauri system SY Cha, CO emission was found at $\approx$ 0.13~au originating from 1570 K gas \citep{Schwarz_2024}. These MIRI observations do not have the required spatial resolution to determine if this gas emission is tracing only the dust depleted inner disk. 

The first detailed physical models of the dust rim were made by \citet{Isella_Natta_2005}. They predicted a natural curved inner rim instead of a vertical wall. Magnetohydrodynamic (MHD) simulations coupled with accretion models incorporated the truncation of the dust depleted inner disk by stellar magnetic fields \citep{Romanova_2003, Bessolaz_2008}. In these models, the gas is lifted out of the disk plane along magnetic field lines and accreted onto the stellar surface, producing hydrogen recombination lines (e.g. Br $\gamma$). Thermally driven or magneto-centrifugal winds carry away angular momentum causing material to accrete onto the star \citep{Ferreira_2006, Suzuki_2010,Bai_2013}.

Observations start to hint more and more at disk emission that originates from the innermost disk. This raises the question how much of the observed line emission originates from the dust depleted inner disk. Proper models including both the dust physics and gas processes are very limited and complex. In this work we do a first step towards bridging the unknown between dust physics and gas chemistry for protoplanetary disks. By creating a physical model that connects both fields we hope to gain a better understanding of the inner region of protoplanetary disks where dust sublimation plays a role. We do this by creating a consistent thermochemical model that includes both the dust depleted inner disk and the classical protoplanetary disk from the sublimation radius onward. To achieve this we add a dust depleted gaseous area to the inner disk of the 2D radiation thermochemical disk model called ProDiMo \citep{Woitke_2009,Kamp_2017,Rab_2018,Thi_2020}. The dust distribution is taken from the output of a multi-dust species magnetohydrostatic disk model \citep{Mario_2025}. This model is based on the magneto-rotational instability (MRI) which provides the underlying dust structure. Subsequently we use ProDiMo to obtain the gas composition, gas temperatures and spectral predictions.

In Section \ref{Method} we explain how the model is constructed, justify our method and introduce the key parameters and elemental abundances that are used. Section \ref{results} analyses the physical structure, temperature profile, chemical network, molecular abundances and presents simulated spectra for the final model including dust depletion. Section \ref{Discussion} will compare our model with previous works, including \ce{H2} formation pathways to discuss model limitations and areas of improvement for future work. 

\section{Model} \label{Method}
This section explains what the most important parameters and assumptions are for how we coupled two codes and generated a consistent model for studying the chemistry in dust depleted inner disk. Subsequently, we describe how we make predictions for molecular emission in the near-IR.

\subsection{ProDiMo}

\begin{figure}[h]
    \centering
    \includegraphics[width=1\linewidth]{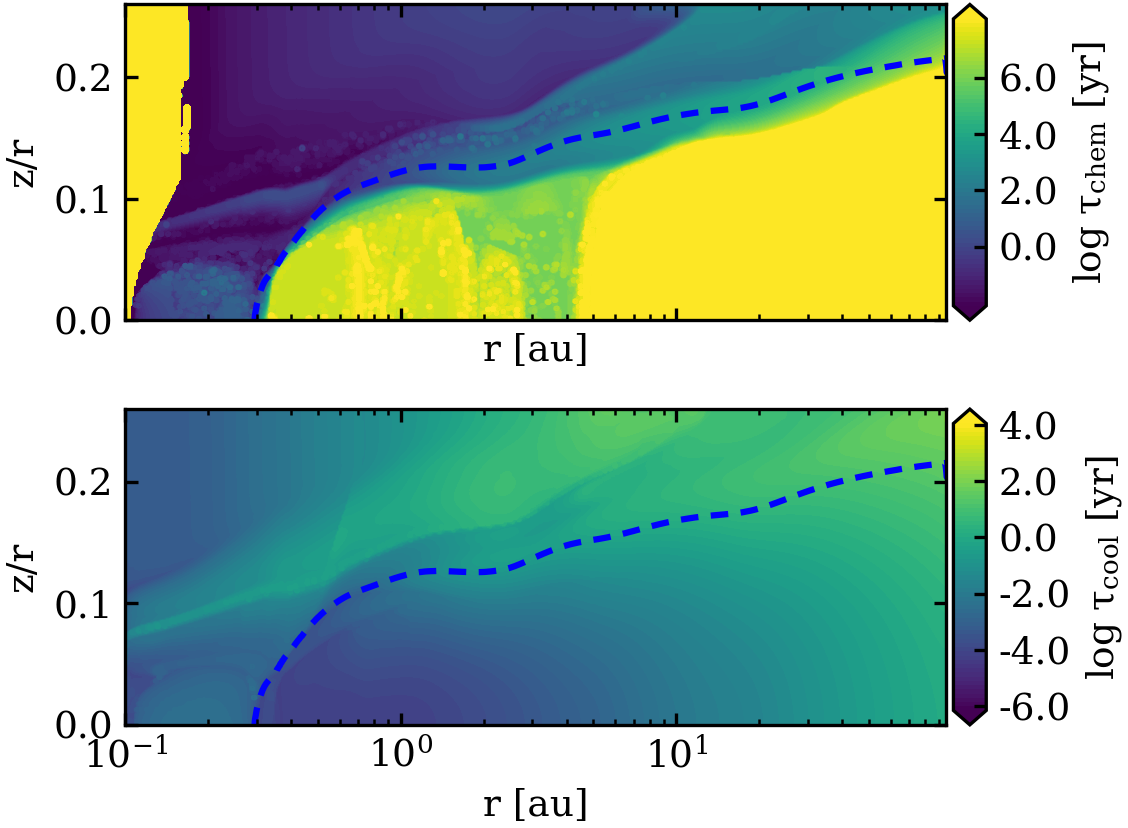}
    \caption{Chemical relaxation timescale $\tau_{chem}$ (top) and cooling relaxation timescale $\tau_{cool}$ (bottom) for our model including a dust free inner environment. The blue dashed line indicates where the visual extinction reaches unity ($A_V = 1$).}
    \label{timescales}
\end{figure}

 ProDiMo\footnote{Version 3, Revision: 605e5c51 2025/08/04}, short for Protoplanetary Disk Model \citep{Woitke_2009,Kamp_2017,Rab_2018,Thi_2020} is a 2D radiation thermochemical disk code that models the physical and chemical structure of protoplanetary disks. In this work we will show 3 different models in a stepwise approach. The first model is the standard Herbig type ProDiMo model. For the second and third model, the physical density setup is based on the output from a 2D multi-dust species magnetohydrostatic model M001 from \citet{Mario_2025}. We use the output from this model as underlying structure in ProDiMo using the interface of \citet[][see section \ref{Setup}]{Rab_2022}. This model contains a more physically realistic density structure for the four dust species corundum (\ce{Al2O3}), iron (Fe), forsterite (\ce{Mg2SiO4}) and enstatite (\ce{MgSiO3}). Then, ProDiMo is used to calculate the gas thermal balance and chemistry until a steady state solution has been found. In the photon-dominated upper layers of a disk, the chemical timescales in the dust depleted inner disk are less than one year (Fig. \ref{timescales}). The chemical relaxation time shows the longest time needed to reach steady state per gridpoint. In the dust depleted region ($<$ 0.4 AU) the chemistry is completely radiation driven, mostly ion-neutral and therefore very fast. The moment dust is abundant the chemistry is dominated by neutral-neutral reactions and takes longer to reach steady state. The average timescale to reach steady state in the observable part of the disk is less than one year. \citet{Zsom_2011} find the radial drift timescale for dust to be on the order of $2 \times 10^4$ years at 1 AU, such that the timescales at 0.1 au are on the order of $\approx 10^2-10^3$ years. Furthermore they  find the viscous timescale to be longer than the radial drift timescale. From this we assume the gas in the inner disk to be in steady state, such that we can run our steady state chemistry model on top of the output from the magnetohydrostatic model. 

\subsection{Dust depleted inner disk setup} \label{Setup}
In order to use the dust density and gas-to-dust ratio of model M001 \citep{Mario_2025} in ProDiMo, we need to make sure that the grid aligns with our ProDiMo grid. The model M001 has a grid of $N_r \times N_{\theta}\  =2304\ \times\ 216$ which is downsized by interpolation in logspace to a grid of size $N_r \times N_{z}\  =500\ \times\ 200$, which is sufficient for calculating the chemistry. Most of our grid points are allocated to the inner few au and the dust rim to make sure our resolution is high enough to resolve the temperature profile of the disk properly. We subsequently use the interpolated dust density and gas-to-dust ratio as input for ProDiMo. Model M001 uses different dust densities for each of the included dust species. In ProDiMo this is simplified  to have a combined dust density where we take the total dust mass fraction and apply it to the whole disk. The magnetohydrostatic model has four dust species with high sublimation temperatures, which are expected to be the main dust component of the dust rim. The mass ratio of these species has been calculated within the magnetohydrostatic model using GGchem \citep{Woitke_2018b, Mario_2025}. The model M001 has a disk extending from 0.1 to 100 au, which we adopted. Table \ref{Parameters_Herbig_Flock}, shows the main parameters used in ProDiMo, where the stellar and dust parameters have been adopted from M001. 

\begin{table}[h]
\begin{center}
\caption{Parameters of ProDiMo.}
\begin{tabular}{lll}
\hline \hline
Parameter & Symbol & Value \\ \hline
\hline
*Stellar mass      & $M_{\star}$      &      2.0 $M_\odot$ \\
*Stellar luminosity  & $L_{\star}$   &      21 $L_\odot$    \\
*Effective temperature    & $T_{\star}$         &  8500 K \\
UV excess & $f_{\text{UV}}$& $ 10^{-4}$\\
UV powerlaw index & $p_{\text{UV}}$ & 3.5 \\
X-ray luminosity  & $L_X$& $10^{29}$ \si{erg.s^{-1}} \\
Interstellar UV & $\chi^{\rm ISM}$ & 1 \\
Cosmic-ray ionization rate & $\zeta_{\rm CR}$ & $1.7\times 10^{-17}$\si{s^{-1}}  \\
Distance & $d$ & 387 pc \\
\hline 
*Disk gas mass & $M_{disk}$ & 0.068 $M_{\odot}$ \\
*Maximum Dust-to-gas ratio & $\delta$  & 0.0046 \\
*Inner disk radius & $R_{\rm in}$& 0.1 au \\
*Tapering-off radius & $R_{\rm tap}$& 100 au \\
\hline
*Dust size       & $a_{\rm min}-a_{\rm max}$  &  0.05-10 \si{\mu m}   \\
Dust size dist. power index & $a_{\rm pow}$ &3.5 \\
*Dust species (volume ratio) &   \ce{Al2O3} & 0.052\\
  & \ce{Fe} &  0.138 \\
  & \ce{Mg2SiO4} &  0.395 \\
  & \ce{MgSiO3} & 0.415 \\
\hline \hline 
\end{tabular}
\label{Parameters_Herbig_Flock}
\tablefoot{The parameters starting with a * are adopted from \citet{Mario_2025}. The remaining parameters are typical values used in ProDiMo models for Herbig disks and explained in detail in \citet{Woitke_2009}.}
\end{center}
\end{table}

\begin{figure}[h]
    \centering
    \includegraphics[width=1\linewidth]{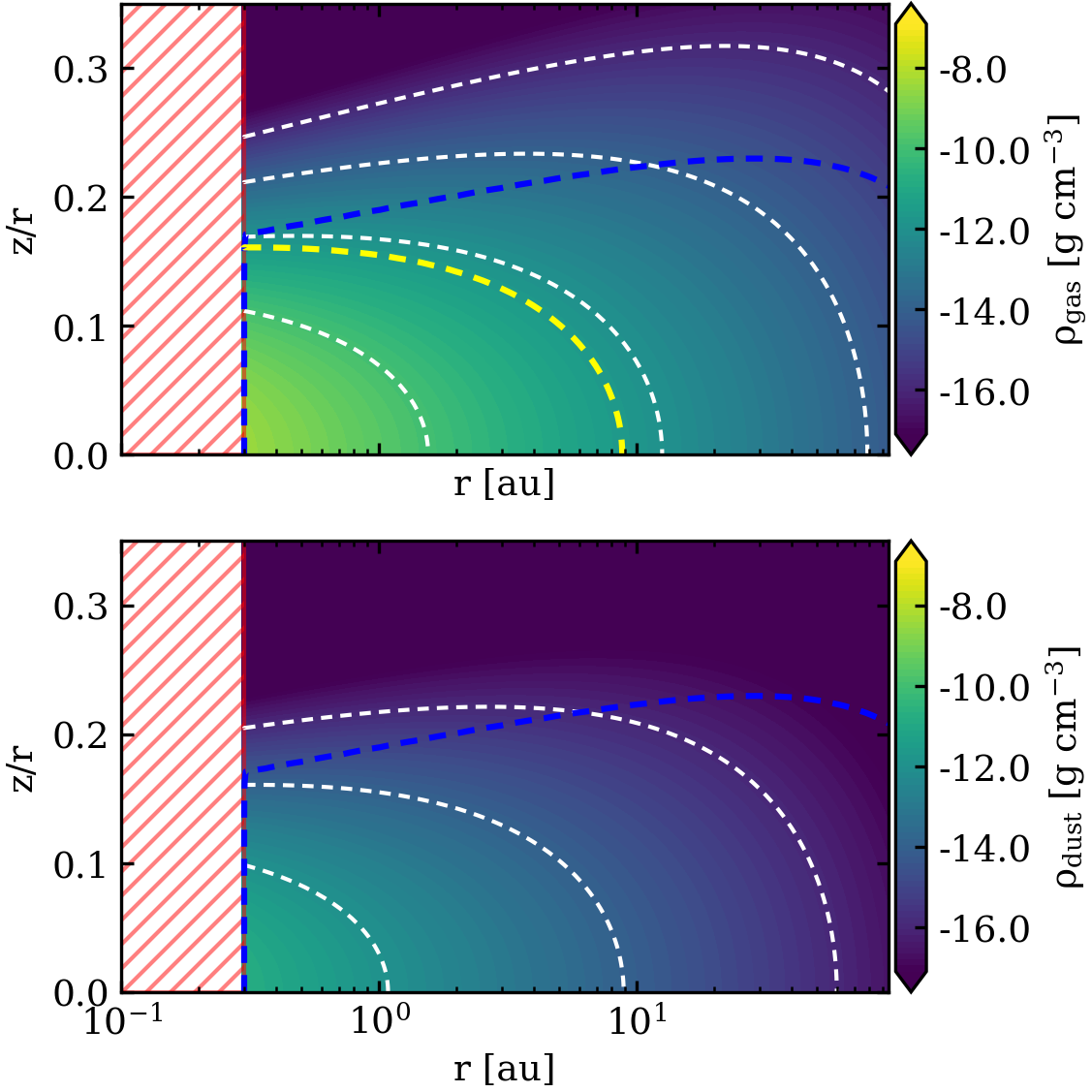}
    \caption{Gas and dust density of a typical ProDiMo Herbig disk model. The red hatched area is generally not modeled as it is beyond the expected dust rim at 0.3 au. The blue dashed line indicates where the visual extinction reaches unity ($A_V = 1$), the white dashed contours align with the values on the colorbar. The yellow dashed line indicates where $n_{\langle H \rangle} = 10^{12}$ \si{cm^{-3}}. }
    \label{M1_dens}
\end{figure}

Figure \ref{M1_dens} shows the standard Herbig star-disk model extending from 0.1 to 100 au with a typical density structure generated using ProDiMo. This model does not extend interior to 0.3 AU, where the dust temperature is high enough that it is consistent with the dust sublimation radius.  The density of the innermost part in this model is not realistic if we were to simply extrapolate the density of dust and gas. The red hatched region has dust depletion but we cannot model this self consistently using only ProDiMo. ProDiMo has been used to model dust depleted regions, such as in transitional disks, or disks with an inner cavity \citep{Woitke_2019}, but not for modeling the gas disk within the dust sublimation radius.

There have been ProDiMo models where this region was constructed trough a rounded rim \citep{Woitke_2024} or a second zone \citep{Woitke_2019}, but our model is the first one created trough proper dust physics. 

\subsection{Chemical network for hot/warm chemistry}
We adapted the large DIANA chemical network with 235 gas and ice species \citep{Kamp_2017}. For this network, we used the reaction rate compilation UMIST2012 \citep{MCElroy_2013} and the additional rates from ChaiTea \citep{Kanwar_2025}. The latter includes 3 body reactions which are very important for the high density ($n_{\langle H \rangle} > 10^{12}$  \si{cm^{-3}}, yellow dashed line in Fig. \ref{M1_dens}) region of the inner disk. \citet{Klarmann_2017} studied the presence of Poly-Aromatic hydrocarbon species (PAHs) or quantum heated particles (QHPs) in the inner region of disks. In their models, the QHP flux does not originate from the dust depleted inner disk, such that PAHs like QHPs do not survive the conditions of the dust depleted inner disk. Hence the PAHs are removed from the network resulting in a total of 228 species. 

\subsection{Elemental Abundances}
Typically in ProDiMo so called metal depleted gas elemental abundances are used \citep{Kamp_2017} to account for elements that are part of the refractory dust distribution. However, as we treat the elemental abundances in a self-consistent way by introducing dust sublimation, Solar abundances from \citet{Asplund_2021} are adopted. The elemental depletion in the gas is based on the dust mass fractions provided by \citet{Mario_2025} (Table \ref{Elemental composition}). In the region where 99\% or more of the dust is sublimated we assume that all elements that are normally locked in dust return to the gas phase resulting in Solar abundances (C/O = 0.59). In the outer disk region, where dust is less than 99\% depleted, we reduce our elemental gas budget with the elements in the dust (Table \ref{Parameters_Herbig_Flock}). From the dust mass fraction, we find that 20\% of our oxygen is locked in dust resulting in a C/O ratio of 0.74 in the outer disk. The selection of dust species does not contain any carbonaceous grains, nor do we have any refractory sulfur species. However, including these for the outer disk would not change our conclusions for the inner disk. Aluminum is not considered here despite corundum being a dust species, as this element is not included in our gas phase chemistry.

\begin{table}[h]
\begin{center}
\caption{Elemental abundances.}
\begin{tabular}{lrrr}
\hline \hline
Element & Outer disk  & $\%$ in dust & Inner disk   \\ \hline
H  & 1 &  &  \\
He & 8.13 $\times 10^{-2}$ &    & \\
C  & 2.88 $\times 10^{-4}$&  & \\
N  & 6.76 $\times 10^{-5}$ & & \\
O  & 3.89 $\times 10^{-4}$ &  20\%  & 4.86 $\times 10^{-4}$  \\
Ne & 1.15 $\times 10^{-4}$ & & \\
Na & 1.66 $\times 10^{-6}$ &  & \\
Mg & 3.55 $\times 10^{-7}$ & 99\% & 3.55 $\times 10^{-5}$ \\
Si & 3.24 $\times 10^{-7}$ &  99\% & 3.24 $\times 10^{-5}$ \\
S  & 1.32 $\times 10^{-5}$ &  & \\
Ar & 2.40 $\times 10^{-6}$ &  &\\
Fe & 4.37 $\times 10^{-6}$ & 85\% & 2.91 $\times 10^{-5}$ \\

\hline \hline
\end{tabular}
\label{Elemental composition}
\tablefoot{Elemental abundances used in the dust depleted inner disk and dusty outer disk in ProDiMo. Solar abundance values are adopted from \citet{Asplund_2021} and given in relative abundance $n_x/n_{\langle H \rangle}$. The outer disks column shows the Solar abundance reduced by the elements locked in dust, and the inner disk column shows the Solar abundance for the 4 species that are affected by dust sublimation.}
\end{center}
\end{table}

\subsection{Molecular line emission and spectra} \label{Esc_prob}
 Beyond the typical molecular data in ProDiMo \citep{Woitke_2024}, we include a selection of lines for \ce{H2O} from the Hitran2020 database \citep{MIKHAILENKO_2020, Gordon_2022} and \ce{SiO} from the ExoMol database \citep{Yurchenko_2022}. This line selection allows us to include many additional lines between 1-5 $\mu$m for \ce{H2O}. The lines in question are both used in gas heating and cooling and to predict observables. Using \ce{H2O} from Hitran2020 instead of the LAMDA database \citep{LAMDA_2005} does limit us to LTE calculations. However as the number density in the dust depleted inner disk is high enough ($n_{\langle H \rangle} \sim 10^{12}-10^{14}$ \si{cm^{-3}}) with respect to the critical density \citep[$n_{\langle H \rangle} \approx 10^{13}$ \si{cm^{-3}} at 5 \si{\mu m},][]{Banzatti_2023}, LTE is a good approximation for \ce{H2O}. We have tested the effect of LTE vs non-LTE treatment of \ce{H2O} in the model and found that the difference in emitted line fluxes is small, but the added high energetic \ce{H2O} lines between 1-5 \si{\mu m} provide a significant amount of gas cooling in the dust depleted inner disk. Hence even though non-LTE is the better treatment for line emission in lower density regions, we think the LTE model is more accurate for the inner disk.

Line predictions in ProDiMo are often made using the escape probability method \citep{Woitke_2018}. Within ProDiMo, the escape probability is also used as an approximate radiative-transfer method to compute the molecular non-LTE level populations of each grid point. The model assumes that a photon produced in a transition has a probability to escape. The escape of this photon regulates the cooling rate. Once the level populations are known for the full grid, ProDiMo uses these together with continuum opacities to create spectra for the whole disk. This method uses the probability for photons to reach the disk surface only in the vertical direction. This method is not exact, as it ignores opacity blending and disk inclination. However it provides a very good estimate for near- and mid-IR spectra at zero degree inclination \citep{Woitke_2018}. Proper observables, including line cubes and predictions for GRAVITY will be in an upcoming paper.

\subsection{Different model setups}
The ProDiMo model with the dust depleted inner disk has been step-wise constructed to ensure that the underlying physics and chemistry is understood. In this paper we will show three different models, that have all been setup using the parameters in Table \ref{Parameters_Herbig_Flock}. 

In the first model (Figure \ref{M1_dens}), the dust and gas density are setup using the standard surface density power law, but the disk is cut at 0.3 au. In the second model, the dust and gas distribution from M001 have been used. For the third model, we use the model setup with the dust and gas distribution from M001 and the elemental abundances in the dust depleted inner disk are enriched to be the Solar values shown in Table \ref{Elemental composition}. 

Using the first two models, we will introduce the physical structure of the dust depleted inner disk and study how the gas temperature changes. The second and third model will be used to illustrate the importance of enhancing the elemental abundances of O, Si, Fe and Mg in the dust depleted region for the spectral profile of the first \ce{SiO} overtone band. For the third and most complete model we will carry out a deeper analysis and study the chemical network, molecular species abundances, vertical temperature profile and provide spectral predictions for the first \ce{CO} overtone and the first \ce{SiO} overtone band. 

\section{Results}\label{results}
\subsection{Disk structure of the inner 1 AU}
\begin{figure}[ht]
    \centering
    \includegraphics[width=1\linewidth]{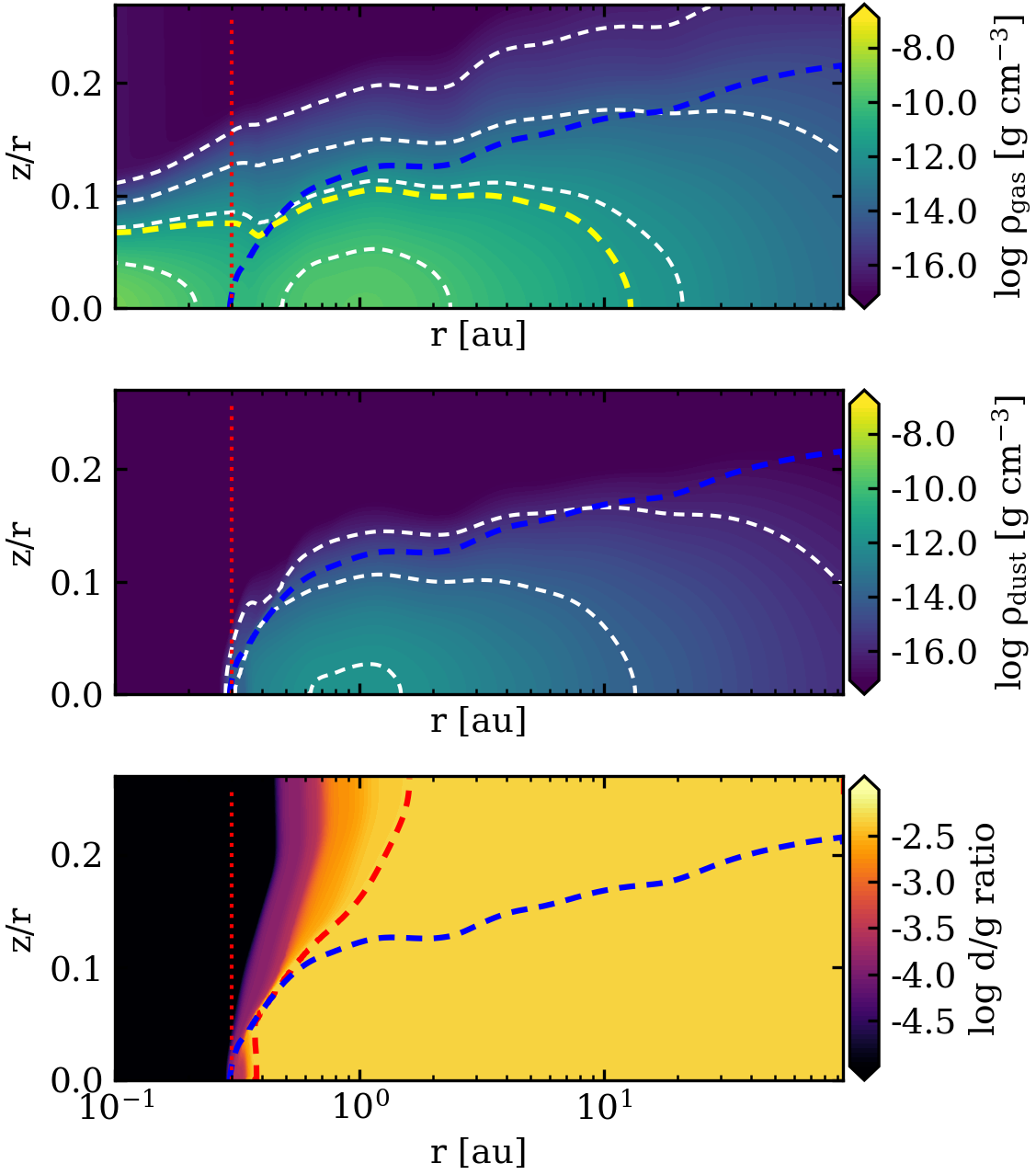}
    \caption{Gas density, dust density and the dust-to-gas ratio of our model with the dust depleted inner disk included. The red vertical dotted line indicates where the standard model is cut. The blue dashed line indicates where the visual extinction reaches unity ($A_V = 1$), the white dashed contours align with the values on the colorbar and the red dashed contour in the bottom panel indicates the outer boundary of the dust depleted inner disk. The yellow dashed line indicates where $n_{\langle H \rangle} = 10^{12}$ \si{cm^{-3}}.}
    \label{M3_dens}
\end{figure}

Previous ProDiMo models in general do not include the small dust-depleted inner region, but instead have the dust sublimation radius as inner radius (see Fig. \ref{M1_dens}). Figure \ref{M3_dens} shows that our new model provides information about the dust free environment between $\sim$0.1--0.35 au in the midplane at z/r = 0. Due to the more physical dust rim shape in the upper disk layers, this region is radially more extended ranging from 0.1 $\sim$ 0.5 au. The dust depleted inner disk is not completely dust free, but has a dust-to-gas ratio at least 3 orders of magnitude lower with respect to the outer disk (third panel of Fig. \ref{M3_dens}).

\begin{figure}[ht]
    \centering
    \includegraphics[width=1\linewidth]{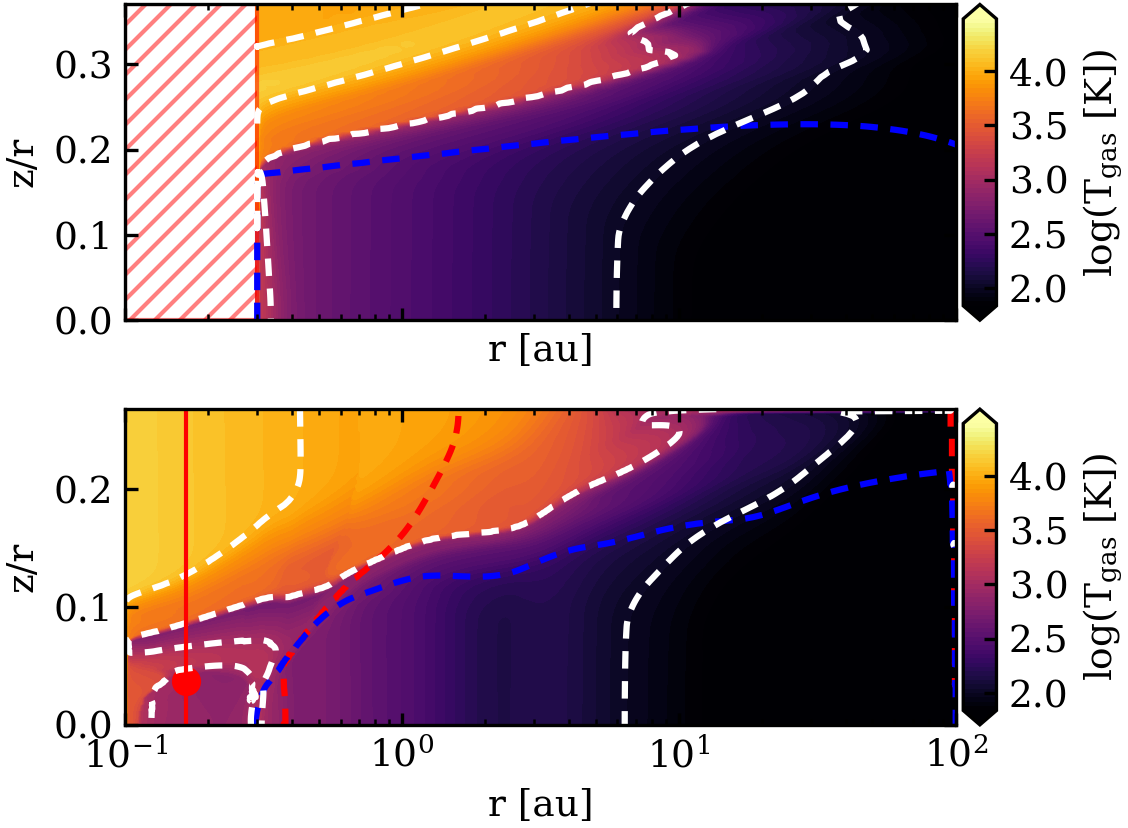}
    \caption{The top panel shows the temperature profile of a standard ProDiMo model and the bottom shows the temperature profile of our model that includes a dust depleted inner disk and increased inner disk gas-phase elemental abundances (see Table \ref{Elemental composition}). The blue dashed line indicates where the visual extinction reaches unity ($A_V = 1$), the white contours show the gas temperatures of 100, 1000 and 10000 K. The red vertical line indicates where the temperature is analyzed in Fig. \ref{Heat_cool_abundance} and the red dot is the grid point at which the chemistry is analyzed (Fig. \ref{Reaction_network}).}
    \label{temp_profile}
\end{figure}

The first and third model are not made to be directly compared, but they do have the same disk mass and stellar parameters. The scale height is different so the upper layers are shifted vertically. However, the density and temperature contours in the midplane for the outer disk (>1 AU) are approximately at the same radius, e.g. the contours in Fig. \ref{M1_dens} and \ref{M3_dens}. Similarly the temperature profile (Fig. \ref{temp_profile}) where the 100 K contour has shifted only by 0.3 au.
 
\subsection{Temperature profile of the inner disk}
Figure \ref{temp_profile} shows that the dust free inner disk (R < 0.3 au) has a complex temperature profile. This pattern originates from the gas heating and cooling balance and is a direct result from the density structure and chemistry. Regions can have similar gas temperatures but completely different dominant heating and cooling processes. The model has 119 heating and 111 cooling processes. 

\begin{figure}[ht]
    \centering
    \includegraphics[width=1\linewidth]{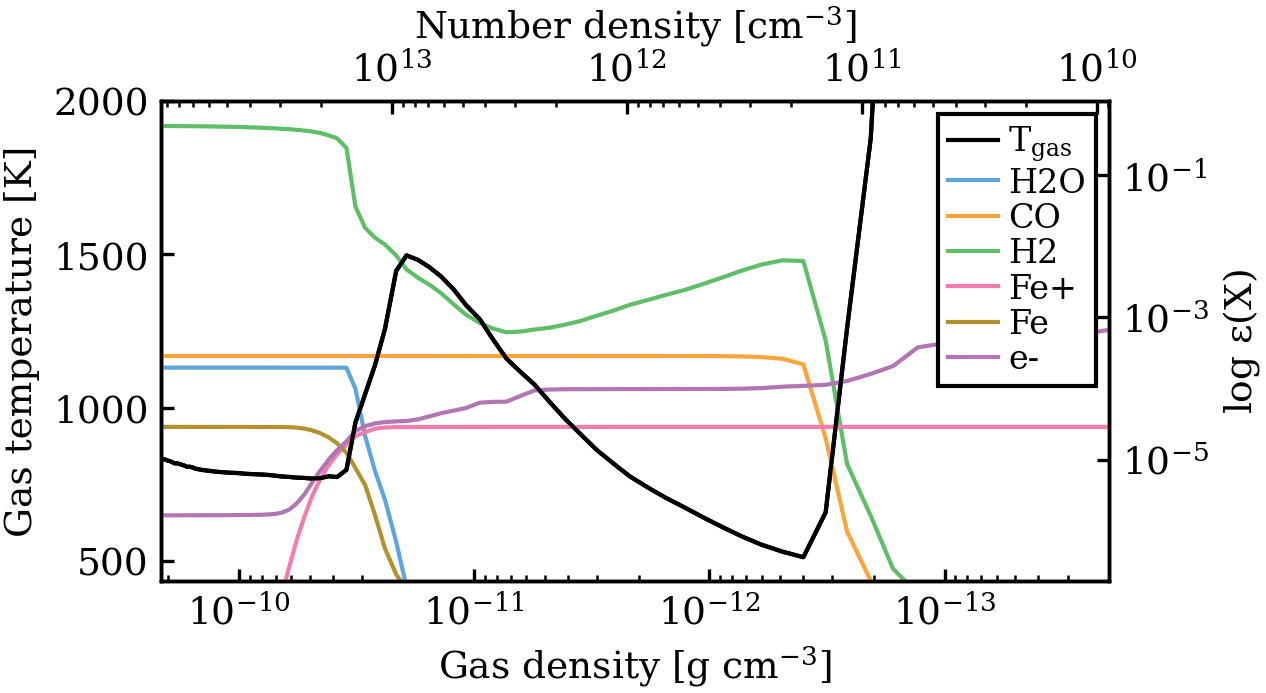}
    \caption{Vertical cut of the disk at r = 0.165 au, showing the gas temperature and molecular abundances with respect to the gas and number density. The left axis indicates the gas temperature and the right axis shows the relative molecular abundance for selected species important to the heating and cooling balance. The midplane is the left y-axis at $\rho_{\text{gas}} \approx 2.15 \times 10^{-10}$ [g \si{cm^{-3}}]. }
    \label{Heat_cool_abundance}
\end{figure}

Figure \ref{Heat_cool_abundance} shows how the gas temperature and several species change as a function of gas density for a vertical cut at 0.165 au (the red vertical line in the bottom panel of Fig. \ref{temp_profile}) with the midplane being on the left hand side. There are three points where the gas temperature profile inverses. The low ($\approx$ 700 K) gas temperature in the midplane of the dust free inner disk is mostly caused by \ce{H2O} line cooling. \ce{H2O} can exist as a consequence of self-shielding and because of the high gas density. The first temperature reversal happens when the gas density drops below, $\rho_{\text{gas}} \approx4 \times\ 10^{-11}$ \si{g\ cm^{-3}}. At this density, Fe is no longer neutralized through collisions with electrons or \ce{H-}. This leads to background heating by FeII lines. Background heating is the radiative heating that is caused by direct excitation through stellar optical and near-IR radiation \citep{Woitke_2009}. Background heating is dominated by stellar photons, as the dust emission is negligible. This heating causes a small ($\approx$ 50 K) gas temperature increase. The lower density also decreases the efficiency of the 3-body reaction forming \ce{H2}. This causes the \ce{H2O} and \ce{H2} abundances to drop, thus decreasing the cooling rate. \ce{H2O} is responsible for $\ge$ 97\% of the total gas cooling inside the dust free inner disk where \ce{H2O} is abundant. 

As the \ce{H2O} abundances decrease, the temperature increases until it reaches a maximum at $\rho_{\text{gas}} \approx2 \times\ 10^{-11}$ \si{g\ cm^{-3}}. The heating and cooling balance here shows two key heating mechanisms; photodissociation heating and chemical heating, are balanced by the rotational and ro-vibrational line cooling of \ce{CO}. Photodissociation heating is linearly and chemical heating quadratic dependent on the gas density. As the gas density decreases, the heating becomes less efficient compared to the linearly dependent cooling. This causes a slow decrease in the gas temperature, until the final temperature reversal is reached at z/r = 0.085 (r,z = 0.165, 0.0149 au), $\rho_{\text{gas}} \approx 3.2\times 10^{-13}$ \si{g\ cm^{-3}}. At higher densities, \ce{CO} is efficiently cycled by forming \ce{HCO} and \ce{CO2} and photo-dissociating back to CO. But if the density drops below $\rho_{\text{gas}} \approx 3.2\times 10^{-13}$ \si{g\ cm^{-3}}, the formation rates decrease leading to lower CO abundances in the surface regions. Having no efficient coolants and being exposed to the stellar and ISM radiation, the gas temperature increases towards the disk surface.

\subsection{Chemical network of the hot inner disk}

\begin{figure}[ht]
    \centering
    \includegraphics[width=1\linewidth]{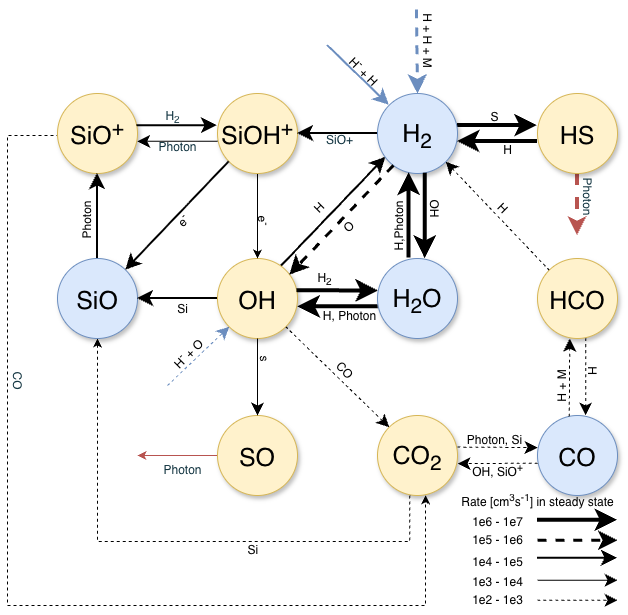}
    \caption{Reaction network of steady state chemistry at r = 0.165 au and z = 0.0061 au where $T_{\text{gas}} = 780\text{K}$. The abundances for the species marked in blue are shown in Fig. \ref{Molecules} and their formation and destruction reactions are described in more detail and shown in Table \ref{H2_formation}, \ref{CO_formation}, \ref{H2O_formation} \& \ref{SiO_formation}.}
    \label{Reaction_network}
\end{figure}

 We analyzed the chemistry in the model with increased elemental abundances for the dust depleted region of the inner disk. We find several abundant molecules besides \ce{CO} that can form at high (> 750 K) temperatures. Abundant \ce{H2} is produced in the absence of dust. The presence of \ce{H2} leads to the formation of many molecules, especially abundant \ce{H2O}. Specifically for this model, we also find high abundances for gaseous \ce{SiO} in the dust depleted inner disk due to the Solar Si abundance. All of these molecular species are connected in steady state. We analyzed a specific grid point that is on the boundary of the hot and cold gas reservoir and has a high abundance of \ce{H2O}. Figure \ref{Reaction_network} shows the chemical reaction network in steady state for r = 0.165 au, z = 0.0061 au, z/r = 0.0371, $T_{\text{gas}} =$ 780 K, $\rho_{\text{gas}} = 6.55\times 10^{-11}$ \si{g\ cm^{-3}}.

\begin{table}[ht]
\caption{Steady state \ce{H2} formation \& destruction pathways in ProDiMo (Fig. \ref{Reaction_network}). $\epsilon$(\ce{H2}) $= 4.47\times 10^{-1}$. M is the combined number density of H, \ce{H2} and He.}
\begin{tabular}{llll}

\hline 
  & Reaction & & k [cm$^{3}$s$^{-1}$]  \\ \hline
  & Formation & & \\
1   & H +  \ce{H2O}            &  $\rightarrow$ OH + \ce{H2}            & $4.80 \times 10^6$     \\
2   & H +  HS                  &  $\rightarrow$ S + \ce{H2}             & $3.71 \times 10^6$     \\
3   & H +  H +  M              &  $\rightarrow$ \ce{H2} + M             & $9.13 \times 10^5$      \\
4   & H +  OH                  &  $\rightarrow$ O + \ce{H2}             & $8.92 \times 10^5$      \\
5   & \ce{H-} +  H             &  $\rightarrow$ \ce{H2} + \ce{e-}       & $5.09 \times 10^4$      \\
6   & \ce{H2O} + photon        &  $\rightarrow$ O + \ce{H2}             & $4.61 \times 10^4$      \\
\\
  & Destruction & & \\
1   & \ce{H2}   + OH           & $\rightarrow$ \ce{H2O}          + H         & $5.07 \times 10^6$  \\
2   & \ce{H2}   + S            & $\rightarrow$ HS                + H         & $4.31 \times 10^6$  \\
3   & \ce{H2}   + O            & $\rightarrow$ OH                + H         & $1.48 \times 10^5$  \\
4   & \ce{H2}   + \ce{SiO+}    & $\rightarrow$ \ce{SiOH+}        + H         & $7.29 \times 10^4$  \\

\hline
\end{tabular}
\label{H2_formation}
\end{table}

 Table \ref{H2_formation} shows the dominant formation and destruction reactions for \ce{H2} in steady state. There are six different pathways to create \ce{H2}. For four of these formation reactions, the reverse rate is higher. This leaves us with only two efficient pathways to create \ce{H2}; the 3-body reaction and the gas‑phase associative detachment reaction (reaction 3 \& 5, Table \ref{H2_formation}). The 3-body reaction is the most important reaction in regions that have a gas density above $\rho_{\text{gas}} \approx 2\times 10^{-11}$ g \si{cm^{-3}} and the associative detachment dominates in less dense regions. The latter cannot sustain high abundances of \ce{H2} (see Fig. \ref{Heat_cool_abundance}). For the 3-body reactions, M is defined as a collective species representing the third body that is approximated by adding the abundances for \ce{H}, \ce{H2} and \ce{He} (M = $n_{\text{H}} + n_{\text{H2}} + n_{\text{He}}$). The relative abundance ($\epsilon (x)$) is defined as the ratio of the number density of the respective species to the number density ($n_{\langle H \rangle}$). $\epsilon (x) = n_x/n_{\langle H \rangle}$ where the number density is calculated as $n_{\langle H \rangle} = n_{\text{H}} + 2 n_{\ce{H2}}$.

\begin{table}[ht]
\caption{Steady state \ce{CO} formation \& destruction pathways in ProDiMo (Fig. \ref{Reaction_network}).  $\epsilon$(\ce{CO)} $=2.88\times 10^{-4}$.}
\begin{tabular}{llll}

\hline 
  & Reaction & & k [cm$^{3}$s$^{-1}$]  \\ \hline
  & Formation & & \\
1   & \ce{CO2} + photon         & $\rightarrow$ CO + O                  & $6.14 \times 10^2$     \\
2   & Si + \ce{CO2}             & $\rightarrow$ SiO + CO                & $1.73 \times 10^2$    \\
3   & H + HCO                   & $\rightarrow$ CO + H2                 & $1.38 \times 10^2$    \\
4   & HCO + photon              & $\rightarrow$ CO + H                  & $2.15 \times 10$    \\
5   & C + OH                    & $\rightarrow$ CO + H                  & $3.99 \times 10^{-2}$    \\
  \\
  % Selected destruction reactions
& Destruction & & \\
1   & OH + CO               & $\rightarrow$ \ce{CO2} + H                 & $6.71 \times 10^2$    \\
2   & H + CO + M            & $\rightarrow$ HCO + M                      & $1.58 \times 10^2$    \\
3   & CO + SiO$^+$             & $\rightarrow$ \ce{CO2} + Si+               & $1.16 \times 10^2$    \\
\hline
\end{tabular}

\label{CO_formation}
\end{table}

\begin{table}[ht]
\caption{Steady state \ce{H2O} formation and destruction pathways in ProDiMo  (Fig. \ref{Reaction_network}). $\epsilon$(\ce{H2O}) $ = 1.98\times 10^{-4}$.}
\begin{tabular}{llll}

\hline 
& Reaction & & k [cm$^{3}$s$^{-1}$]  \\ \hline
& Formation  &   & \\
1 & \ce{H2}  + OH       & $\rightarrow$ \ce{H2O}  + H   & $4.31\times 10^6$ \\
\\
& Destruction & & \\
1 & H + \ce{H2O}        & $\rightarrow$ OH + \ce{H2}    & $3.71\times 10^6$ \\
2 & \ce{H2O}  + photon  & $\rightarrow$ OH + H          & $5.56\times 10^5$ \\
3 & \ce{H2O}  + photon  & $\rightarrow$ O + \ce{H2}     & $4.61\times 10^4$\\
\hline
\end{tabular}
\label{H2O_formation}
\end{table}

\begin{table}[ht]
\caption{Steady state \ce{SiO} formation \& destruction pathways in ProDiMo (Fig. \ref{Reaction_network}). $\epsilon$(\ce{SiO}) $=4.53\times 10^{-8}$.}
\begin{tabular}{llll}

\hline 
& Reaction & & k [cm$^{3}$s$^{-1}$]  \\ \hline
    & Formation         &   & \\
1   & OH + Si           & $\rightarrow$ SiO + H  & $3.36\times 10^4$ \\
2   & SiOH$^+$ + e$^-$     & $\rightarrow$ SiO + H  & $3.35\times 10^4$ \\
3   & Si + \ce{CO2}          & $\rightarrow$ SiO + CO & $1.73\times 10^2$ \\
 \\ 
& Destruction & & \\
1 & SiO + photon & $\rightarrow$ SiO$^+$ + e$^-$ & $6.72\times 10^4$ \\
\hline
\end{tabular}

\label{SiO_formation}
\end{table}

\ce{CO}, \ce{H2O} and \ce{SiO} are almost never chemically destroyed to atoms. Figure \ref{Reaction_network} shows that they are all connected. Besides the atoms making up these species (H, C, O, Si), only two molecular species are needed to chemically form these three molecules: \ce{H2} and \ce{OH}. At the selected grid point, \ce{SiO} is not stable and has a low ($5 \times 10^{-8}$) relative abundance. At lower temperatures (< 900 K), oxygen gets preferentially incorporated into \ce{H2O} instead of \ce{SiO}. To have abundant \ce{SiO} we need a destruction of \ce{H2O}.

\begin{figure}[ht]
    \centering
    \includegraphics[width=1\linewidth]{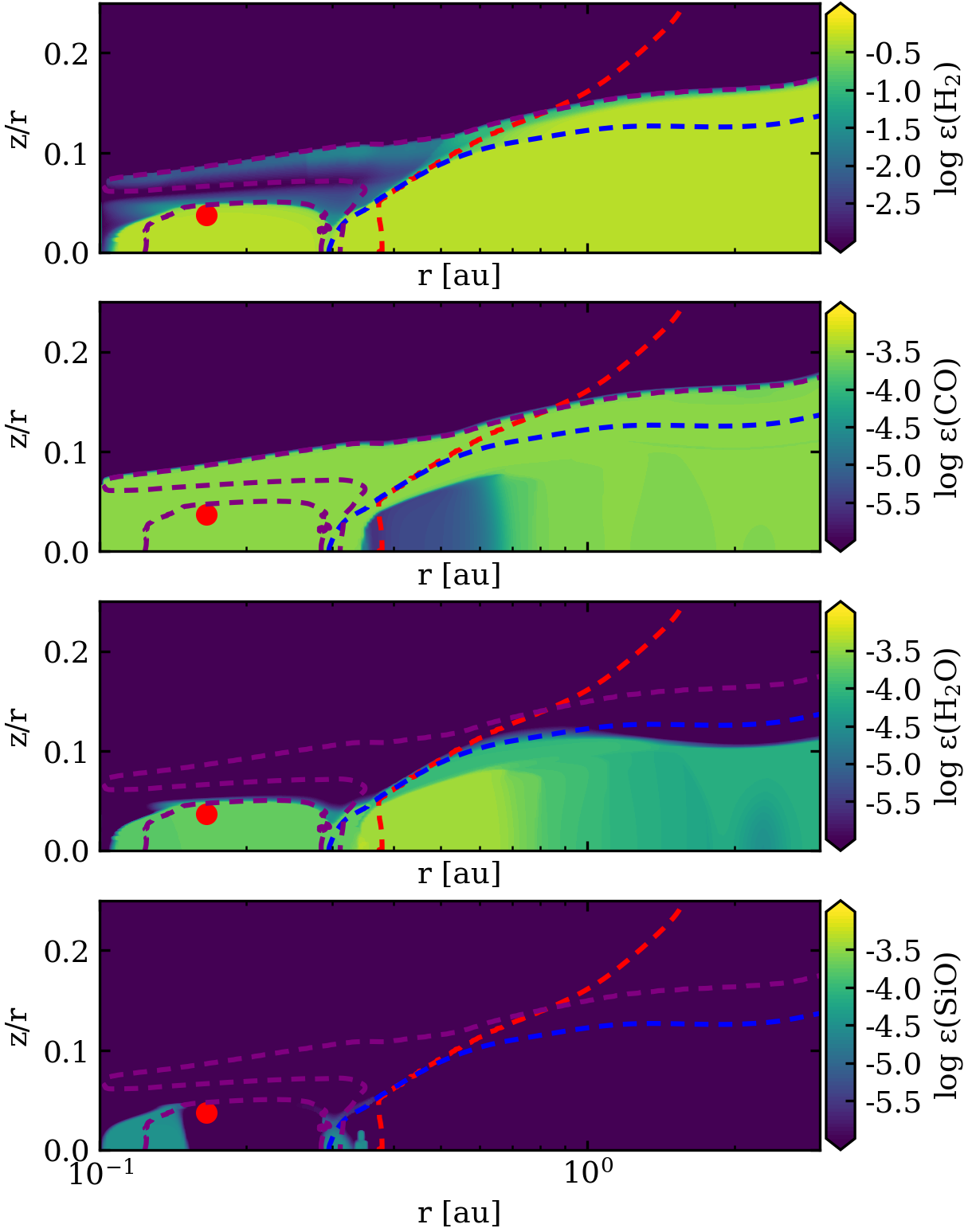}
    \caption{Abundances of \ce{H2}, \ce{CO}, \ce{H2O} and \ce{SiO} for the model with solar abundance gas phase abundances in the inner disk. The purple contour shows where the gas temperature is 1000 K, the red contour shows where the dusty disk starts and the blue dashed line indicates where the visual extinction reaches unity ($A_V = 1$). The red dot indicates the grid point at which the chemistry has been analyzed.}
    \label{Molecules}
\end{figure}

Between 0.4 and 0.7 au, the midplane shows a CO poor region (second panel Fig. \ref{Molecules}). This is an artifact from running steady state chemistry. However, this region is below the $A_v$ = 1 line and therefore not observable and not impacting any results shown in the paper.

\subsection{Elemental enrichment of the dust depleted inner disk}
 The elemental enrichment of \ce{O}, \ce{Mg}, \ce{Si} and \ce{Fe} causes changes to the chemistry and heating-cooling balance. The enrichment changes the gas temperature in the water dominated region by less than 5$\%$ and the gas temperature in the warm CO layer by less than 10$\%$. However, the increase in elemental abundance for \ce{Si} results in about two orders of magnitude higher SiO gas abundances in the dust depleted inner disk (see Fig. \ref{Molecules}). 
 
 \subsection{Near-IR spectral predictions}
\begin{figure}[ht]
    \centering
    \includegraphics[width=1\linewidth]{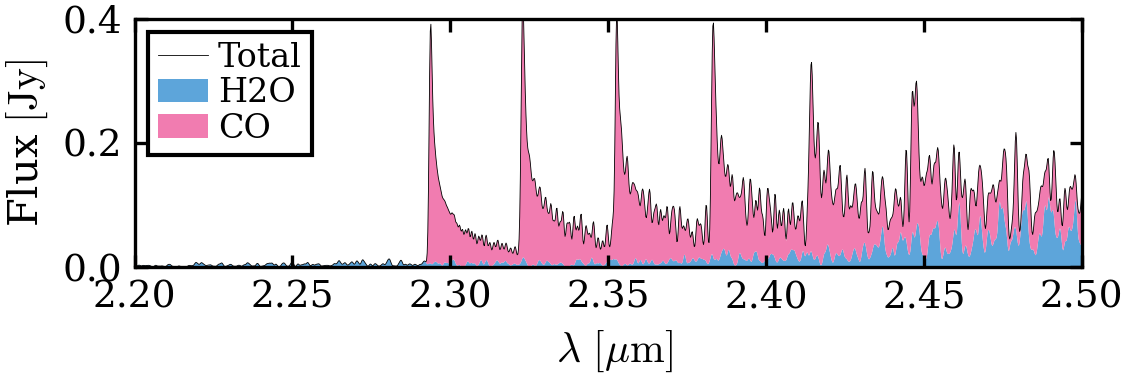}
    \caption{Escape probability spectra showing the first CO overtone lines. This model includes dust depletion and elemental enrichment in the inner disk and is convolved using a spectral resolving power of 3000.}
    \label{CO_overtone}
\end{figure}

 The elemental enrichment of Si has a strong effect on the near-IR emission spectrum for the first \ce{SiO} overtone. Using the escape probability method, we calculate near-IR emission spectra where the object is assumed to be at a distance of 387 pc. We chose this distance to put our predictions in context to JWST observations by \citet{Kaufer_2025}. However, we emphasize that our model was not adapted to match the properties of that disk. We simulate the first CO overtone emission using the escape probability between 2.2 and 2.5 \si{\mu m} (Fig. \ref{CO_overtone}).

\begin{figure}[ht]
    \centering
    \includegraphics[width=1\linewidth]{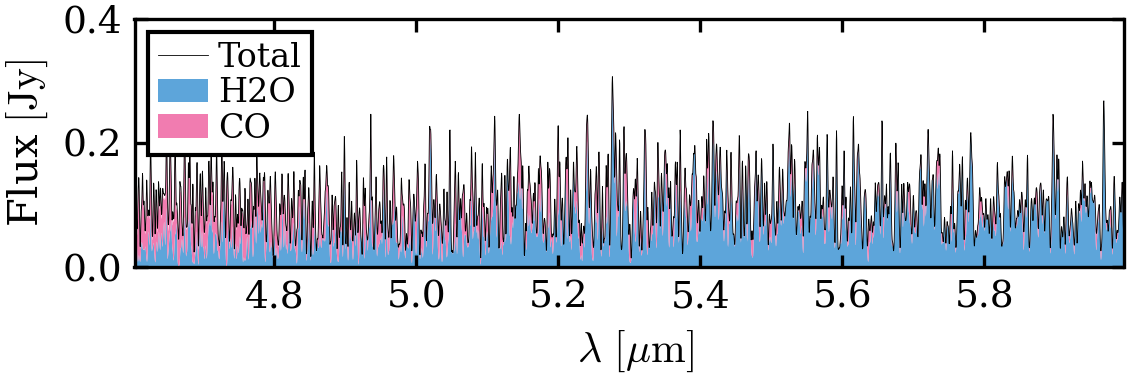}
    \caption{Escape probability spectra showing the \ce{H2O} and \ce{CO} lines between 4.6 and 6 \si{\mu m}. This model includes dust depletion and elemental enrichment in the inner disk and is convolved using a spectral resolving power of 3000.}
    \label{H2O_CO}
\end{figure}

As a result of including all water lines we create a quasi continuum of water around 5 \si{\mu m}. Figure \ref{H2O_CO} shows how dense the water lines are populated. As mentioned before, the escape probability spectra do not include the disk rotational broadening or opacity overlap. If these additional effects are taken into account, the quasi continuum will be even stronger. 

\begin{figure*}[ht]
    \centering
    \includegraphics[width=1\linewidth]{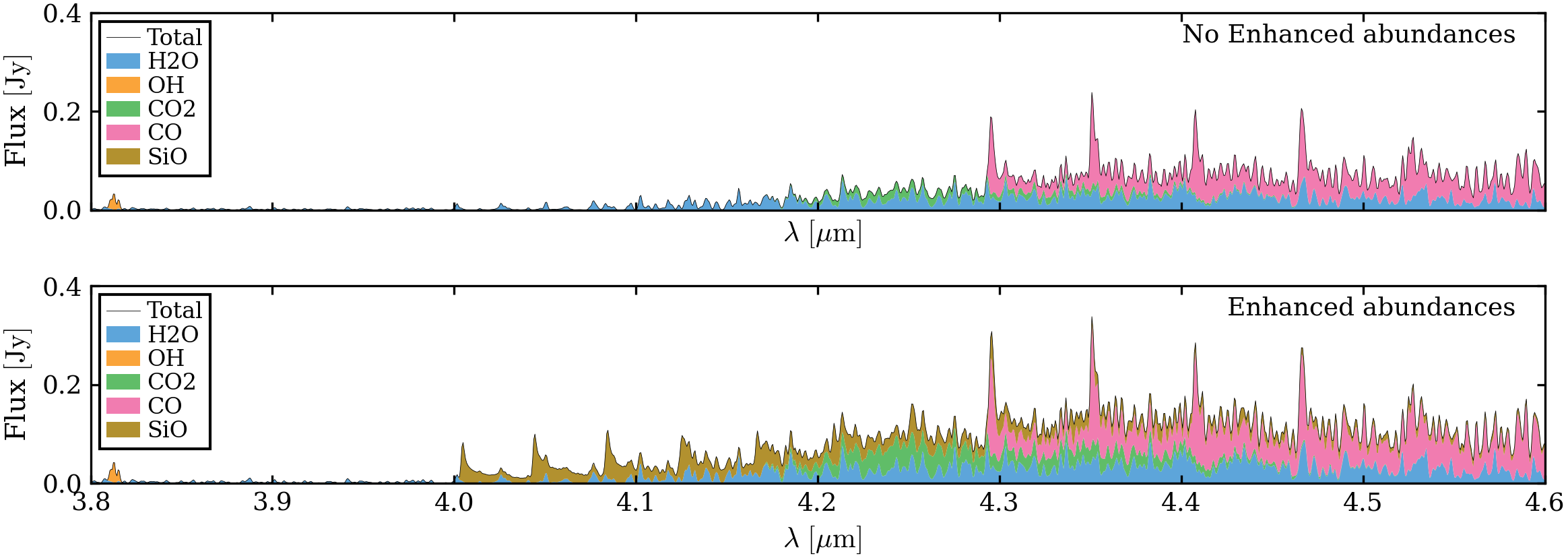}
    \caption{Escape probability spectra from our disk model showing the first SiO overtone lines originating from the dust depleted inner disk of a model with and without elemental enhancement convolved using a spectral resolving power of 3000. The black line shows the combined spectra of molecules in the model and the colors indicate how much of the flux originates from specific molecules.}
    \label{sneakpeak2}
\end{figure*}

We also simulate the first \ce{SiO} overtone emission between 3.8 and 4.6 \si{\mu m}, along with CO, \ce{H2O}, OH and \ce{CO2} for both a model where the elements in the inner have been enhanced and one where these are not enhanced (Fig. \ref{sneakpeak2}). We find that all of our \ce{SiO} overtone emission depends on the enhancement of the elemental abundances, while the \ce{CO} overtone is hardly affected by the difference in elemental abundance. Next to the \ce{SiO} overtone, we find four very strong peaks for high J (J > 60) ro-vibrational \ce{CO} line transitions of the fundamental band. These emission feature can be seen in Fig. \ref{sneakpeak2} between 4.3-4.6 $\mu$m. We will elaborate more on them in the discussion.

\section{Discussion}\label{Discussion}
\subsection{Comparing to previous ProDiMo models}
In typical ProDiMo models \citep[e.g,.][]{Kamp_2023, Woitke_2024} it is common to enhance the gas density such that the dust-to-gass mass ratio (d/g) = 0.001 for the inner 10 au. This adjustment was needed to match the observed Spitzer mid-IR molecular line strengths \citep{Meijerink_2009, Woitke_2018}. We find that at least 90\% of the line emission for \ce{CO} and \ce{H2O} between 1 and 28 $\mu$m originates from the dust depleted inner disk within 0.3 au. A d/g ratio of 0.001 can be obtained naturally in disk models that include full dust evolution and settling \citep{Greenwood_2019}. In our self-consistent models, the d/g ratio is $\approx 10^{-10} $ in the dust depleted inner disk, while beyond 0.4 au the d/g ratio is 0.0046. The canonical d/g ratio is 0.01, but as a result of the underlying MHD model we only include a narrow range of dust grains sizes excluding part of the dust reservoir. However, using only this range is fine for this work as the continuum flux in the near-to mid IR wavelength range is regulated by 0.1 to 5 \si{\mu m} size grains \citep{Jang_2025}.

\subsection{Comparison to existing observations}
\citet{Kaufer_2025}, observed the Herbig star HD35929 as part of the MINDS program \citep{Kamp_2023, Henning_2024}. They find strong gas phase molecular line emission for \ce{SiO} between 8 and 8.7 \si{\mu m}. They find molecular column densities that suggest the disk has a low dust opacity and/or a low d/g ratio. Many of their emission lines are spectrally resolved, placing the gas emission near the star at a typical distance of 0.1-0.2 au. They also detect mid-IR water emission and derived a temperature of $\approx 850$ K. We find that our model predicts the fundamental \ce{SiO} emission to be a factor 2-3 and our \ce{H2O} emission between 5 and 7 micron about a factor 5-7 stronger than what they observe. Our modeled water spectrum is so densely populated with lines that it produces a pseudo continuum when convolved to MIRI/MRS spectral resolution. We also predict that the first overtone band of SiO (4-4.3 \si{\mu m}) has a similar flux strength as the fundamental band. This could be observed either with JWST/NIRSPEC or from the ground with VLTI/CRIRES. There is not yet a detection of the first SiO overtone for Herbig type disks in the literature; it has only been detected in disks around massive evolved stars \citep{Kraus_2015}.

\subsection{Condensation and sublimation} \label{Condensate}
Figure \ref{temp_profile} shows that inside the dust depleted inner disk the gas cools down to below 700 K (T$_{gas}$ = 680 K at r = 0.22 au, z/r = 0.04). At these temperatures, we expect most dust species to re-condensate. The hypothetical region in which dust can exist both in the gaseous and solid state is often called the thermostat region \citep{Min_2011,McClure_2025}. An open question for this dust depleted region is wether the temperatures to sublimate and condensate dust are the same or even similar. In an environment with a low d/g ratio and the presence of intense radiation fields dust might sublimate at lower temperatures compared to an environment where d/g = 0.01. 

\subsection{Temperature jump with respect to the radial hydrogen column density}
\begin{figure}[ht]
    \centering
    \includegraphics[width=1\linewidth]{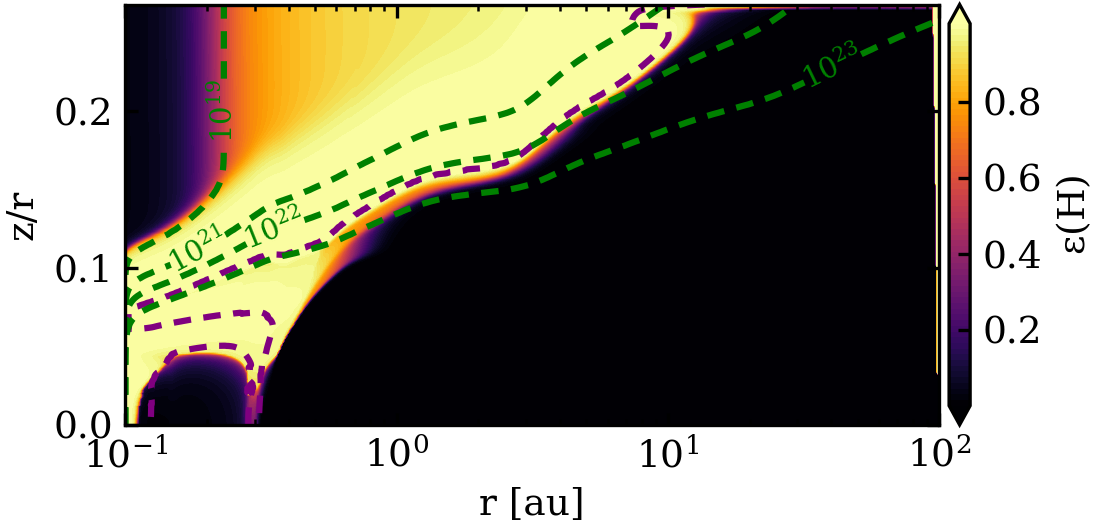}
    \caption{Atomic hydrogen abundance, with the 1000 K temperature contour in purple. In green are the radial hydrogen column density contours for $10^{19}, 10^{21}, 10^{22} \text{ and } 10^{23}$ \si{cm^{-2}}}.
    \label{temp_jump}
\end{figure}

An important region in protoplanetary disk modeling is the transition between the ionized, UV-irradiated disk atmosphere and the UV-shielded midplane layers. Simplified chemistry models coupled with hydrodynamics predict atomic hydrogen column densities of N$_{\langle H \rangle} \approx10^{19}$ \si{cm^{-2}} \citep{Nakatani_2018a, Nakatani_2018b}, which marks the extreme-UV absorption layer and H$^{+}$-H transition. \citet{Flores_2020} mark the X-Ray/far-UV absorption layer at a radial hydrogen column density of  N$_{\langle H \rangle} = 10^{22}$ \si{cm^{-2}}. This value is only valid for radii larger than 1 au as it depends on the presence of dust. In our models, the latter one defines the important transition in which the gas temperature increases above 1000 K in nearly the whole disk atmosphere from 1 to 10 au as seen in Fig. \ref{temp_profile}. We calculated and confirmed that in our thermochemical model a radial hydrogen column density N$_{\langle H \rangle} \approx10^{19}$ \si{cm^{-2}} lies very close to the H$^{+}$-H transition. For the H-\ce{H2} transition however we do not find the transition exactly at N$_{\langle H \rangle} = 10^{22}$\si{cm^{-2}}, but we find a range for the radial column density that changes with radius. Figure \ref{temp_jump} shows how this transition changes with respect to the radial column density. We find between 1-10 au that the transition changes from a radial column density of N$_{\langle H \rangle} \approx 8 \times 10^{22}$\si{cm^{-2}} at 1 au to N$_{\langle H \rangle} \approx 5 \times 10^{21}$\si{cm^{-2}} at 10 au.

\subsection{Is the MRI assumption valid for the whole inner disk?}
The magnetohydrostatic model we use as foundation for the dust and gas density is dependent on the gas temperature. Using ProDiMo, we recalculate the gas temperature including the chemistry. This gives us a region with low ($\approx$ 700 K) temperatures in part of the inner disk contrary to what the hydrostatic model finds. Our temperatures are low enough for a small part of the Fe in the gas to be neutral as can be seen in Fig. \ref{Heat_cool_abundance}. MRI relies on the magnetic fields being coupled to disk material and this requires ionized gas. Our midplane gas temperatures can perhaps even lead to dust re-condensation (see sect. \ref{Condensate}). Both of these effects can suppress MRI-driven turbulence and non-ideal MHD effects \citep{Lesur_Flock_2023}. In order to properly test if the MRI assumptions holds with respect to chemical cooling timescales, we need to run the gas chemistry inside the magnetohydrostatic model.

\subsection{High J-level CO emission lines}
Figure \ref{sneakpeak2} shows very strong CO emission peaks between 4.3 and 4.6 \si{\mu m}. This emission originates from high (>60) J-level line transitions. In the model the molecular data for higher level rotational states are calculated by extrapolation \citep{Thi_2013}. In the future, the validity of these extrapolations needs to be re-visited.

\subsection{\ce{H2} formation in a dust depleted environment}
Studies on \ce{H2} formation in high temperature and dust poor environments are limited. \citet{Thi_2005_H2} proposed three reaction pathways to produce \ce{H2} in the absence of dust. All three are included in our model. Their 3-body reaction, H + H + H $\rightarrow$ \ce{H2} + H is included as H + H + M $\rightarrow$ \ce{H2} + M. Where M is the combined density of H, \ce{H2} and He. They find that the 3-body reaction becomes important at hydrogen densities above $n_{\langle H \rangle} \approx 10^{12}$ \si{cm^{-3}}. \citet{Kanwar_2025} found that to produce abundant \ce{H2} trough 3-body reactions a density of $n_{\langle H \rangle} \approx 10^{12}$ \si{cm^{-3}} is not enough and we need at least $n_{\langle H \rangle} \approx 10^{13}$ \si{cm^{-3}}. We confirm that in our model we need at least $n_{\langle H \rangle} \approx 10^{13}$ \si{cm^{-3}} (see Fig. \ref{Heat_cool_abundance}) to turn H molecular.

\citet{Sternberg_2021} investigated the formation of \ce{H2} in cold (100 K) dust free gas. All of their formation pathways are included in the existing ProDiMo network. Other studies on dust free \ce{H2} formation focus mostly on low density primordial dust free gas, where \ce{H2} is predominantly produced trough the gas‑phase associative detachment reaction, \ce{H-} + H $\rightarrow$ \ce{H2} + \ce{e-}. Even though our model does not have any regions of cold dust-free gas, this reaction is also important at higher temperatures as it is our second most important \ce{H2} formation reaction aside the 3-body reaction.

\subsection{CO overtone emission}
In Fig. \ref{CO_overtone} we show that our model produces CO overtone emission. Because GRAVITY is an interferometric instrument which can only observe relative flux, there is no direct comparison we can do. However, the object for which we compared our fundamental mid-IR \ce{SiO} fluxes has also been observed with X-shooter. \citet{Ilee_2014} find that the relative flux of the first CO overtone bandhead in HD35929 is roughly 10$\%$ of the continuum. Our scaled model shows a CO overtone peak flux of $\approx 15 \%$ relative to the continuum. This is a very good agreement given that we did not aim to fit this object. Next to the CO overtone emission, we also produce some weak water lines in this wavelength region. Such weak water lines have been observed only once in 51Oph \citep{Thi_2005}. In future studies, we will use this model to create interferometric observables for GRAVITY inserting linecubes from the model into a GRAVITY simulator. 

\subsection{Gas opacity}
The underlying magnetohydrostatic model (M001) by \citet{Mario_2025} does include gas opacity to calculate the location of the inner dust rim. However the value used $\kappa_{gas} = 10^{-6}$ \si{cm^2 /g}, is on the low side and considered to be an average gray opacity that is constant across temperature and wavelength. A recent work by \citet{cecil_2026} used the same modeling code as \citet{Mario_2025}, and they show how increasing the gas opacity will move the dust rim inward. This shift of the dust rim will likely lower the expected line emission. However for the current model, the K-band half-light radius matches very well with GRAVITY YSO observations \citep{Mario_2025}. Hence we do expect the dust rim to be around 0.35 au for this type of disk. 

Currently the gas opacity is not included in the dust radiative transfer for our ProDiMo model. This is why our visual extinction ($A_V$) = 1 reaches the midplane when dust sublimates (see Fig. \ref{M3_dens}). However, gas opacity effects are treated within the chemistry through self-shielding for the most important molecules \citep{Woitke_2024}. This shielding effect is also considered for the heating and cooling balance. The predicted spectra account for the gas optical depth based on the specific molecular vertical gas column density.

\subsection{Modeling limitations and future improvements}
The model has several technical limitations, that we list below: 
\begin{itemize}
    \item Many of the chemical rates we used in our model are uncertain as these have not been measured for high temperatures. In future works, we plan to investigate if including combustion chemistry or 3-body reactions for larger molecules are important for the chemical inventory of the dust depleted inner disk. 
    \item We use self-shielding for all molecules on itself. However, with the exception of \ce{C}, \ce{H2} and \ce{CO}, species cannot shield different molecular species, a process also called alien shielding. In the radiation rich environment that we are interested in, the shielding is very important to keep molecules from being photodissociated. Thus future studies should include alien shielding for more molecules. 
    \item The inner radius of our model is set at 0.1 au. This is not for a physical reason, but a direct consequence that the underlying magnetohydrostatic model stops here. For future models, we would like to include the disk up to the co-rotation radius. The effect on the predicted line fluxes should be small as the emitting surface area is at most 12.5 $\%$ larger in the most extreme scenario, where the inner radius extends all the way to the stellar surface. 
\end{itemize}

\section{Conclusion} \label{conclusion}
In this work, we have made the first steps towards accurately modeling the inner regions of protoplanetary disk including a realistic dust distribution based on dust sublimation. The presented model combines a magnetohydrostatic simulation with a radiation thermochemical disk model to include both dust physics and gas chemistry. Our main results are the following: 

\begin{itemize}
    \item The dust depleted inner disk gas is chemically rich, showing large abundances of \ce{H2}, \ce{CO}, \ce{H2O} and \ce{SiO} which are the result of a self consistent chemical model. 
    \item The inner disk has a complex temperature profile. In the absence of dust, the heating and cooling of the gas is dominated by molecular lines, such that the chemistry becomes crucial. We find that \ce{H2O} is a very efficient coolant in the dense midplane region, cooling it down to $\approx$ 700 K. 
    \item We find that the dust depleted inner disk (0.1-0.3 AU) in our model is responsible for at least 90\% of the emitted line fluxes for \ce{CO} and \ce{H2O} between 1 and 28 \si{\mu m}. 
    \item A large reservoir of abundant gas phase \ce{H2O} ($\epsilon ({\ce{H2O}}) = 1.98\times 10^{-4}$) is present in the dust depleted inner disk. This hot \ce{H2O} produces a pseudo continuum in the spectrum between 4.6 to 6 $\mu$m when convolved to MIRI/MRS spectral resolution. Taking the possibility of this into account is important to properly fit observations.
    \item In our self consistent inner thermochemical disk model, the gas temperatures are high enough to produce the commonly observed first CO overtone line emission. 
    \item We reproduce the recently observed \ce{SiO} mid-IR emission reported by \citet{Kaufer_2025}, if we increase the elemental abundances to include the material from sublimated dust. This suggests that observing the fundamental and/or first overtone lines of \ce{SiO} are strong indications of a dust free inner disk, where Si has been returned to the gas phase.
\end{itemize}

In future studies, we will focus on modeling specific Herbig type disks and creating proper synthetic observables for GRAVITY+ and JWST. We will also expand our model to T~Tauri type disks. 

\begin{acknowledgements}
The authors are grateful to the referee for a constructive and positive report that improved the paper.

CHR acknowledges the support of the Deutsche Forschungsge-meinschaft (DFG, German Research Foundation) Research Unit “Transition discs”—325594231.

\end{acknowledgements}

\bibliographystyle{aa.bst} 
\bibliography{References.bib} 

\end{document}